\documentclass[conference]{IEEEtran}
\IEEEoverridecommandlockouts
\usepackage{cite}
\usepackage{amsmath,amssymb,amsfonts}
\usepackage{algorithmic}
\usepackage{graphicx}
\usepackage{textcomp}
\usepackage{xcolor}
\usepackage{booktabs}
\usepackage{amssymb}
\usepackage{pifont}
\newcommand{\cmark}{\ding{51}}%
\newcommand{\xmark}{\ding{55}}%

\def\BibTeX{{\rm B\kern-.05em{\sc i\kern-.025em b}\kern-.08em
    T\kern-.1667em\lower.7ex\hbox{E}\kern-.125emX}}
\begin{document}

\title{On Frequency-Wise Normalizations \\ for Better Recording Device Generalization \\ in Audio Spectrogram Transformers}
\author{\IEEEauthorblockN{Paul Primus$^1$, Gerhard Widmer$^{1,2}$}
\IEEEauthorblockA{$^1$Institute of Computational Perception (CP-JKU)  \\
$^2$LIT Artificial Intelligence Lab\\
Johannes Kepler University, Austria}
}

\maketitle

\begin{abstract}
Varying conditions between the data seen at training and at application time remain a major challenge for machine learning. 
We study this problem in the context of Acoustic Scene Classification (ASC) with mismatching recording devices. 
Previous works successfully employed frequency-wise normalization of inputs and hidden layer activations in convolutional neural networks to reduce the recording device discrepancy. 
The main objective of this work was to adopt frequency-wise normalization for Audio Spectrogram Transformers (ASTs), which have recently become the dominant model architecture in ASC.
To this end, we first investigate how recording device characteristics are encoded in the hidden layer activations of ASTs. 
We find that recording device information is initially encoded in the frequency dimension; however, after the first self-attention block, it is largely transformed into the token dimension. 
Based on this observation, we conjecture that suppressing recording device characteristics in the input spectrogram is the most effective. 
We propose a frequency-centering operation for spectrograms that improves the ASC performance on unseen recording devices on average by up to $18.2$ percentage points.

\end{abstract}


\section{Introduction}
\begin{figure}[t]
  \centering
    \includegraphics[width=1\linewidth]{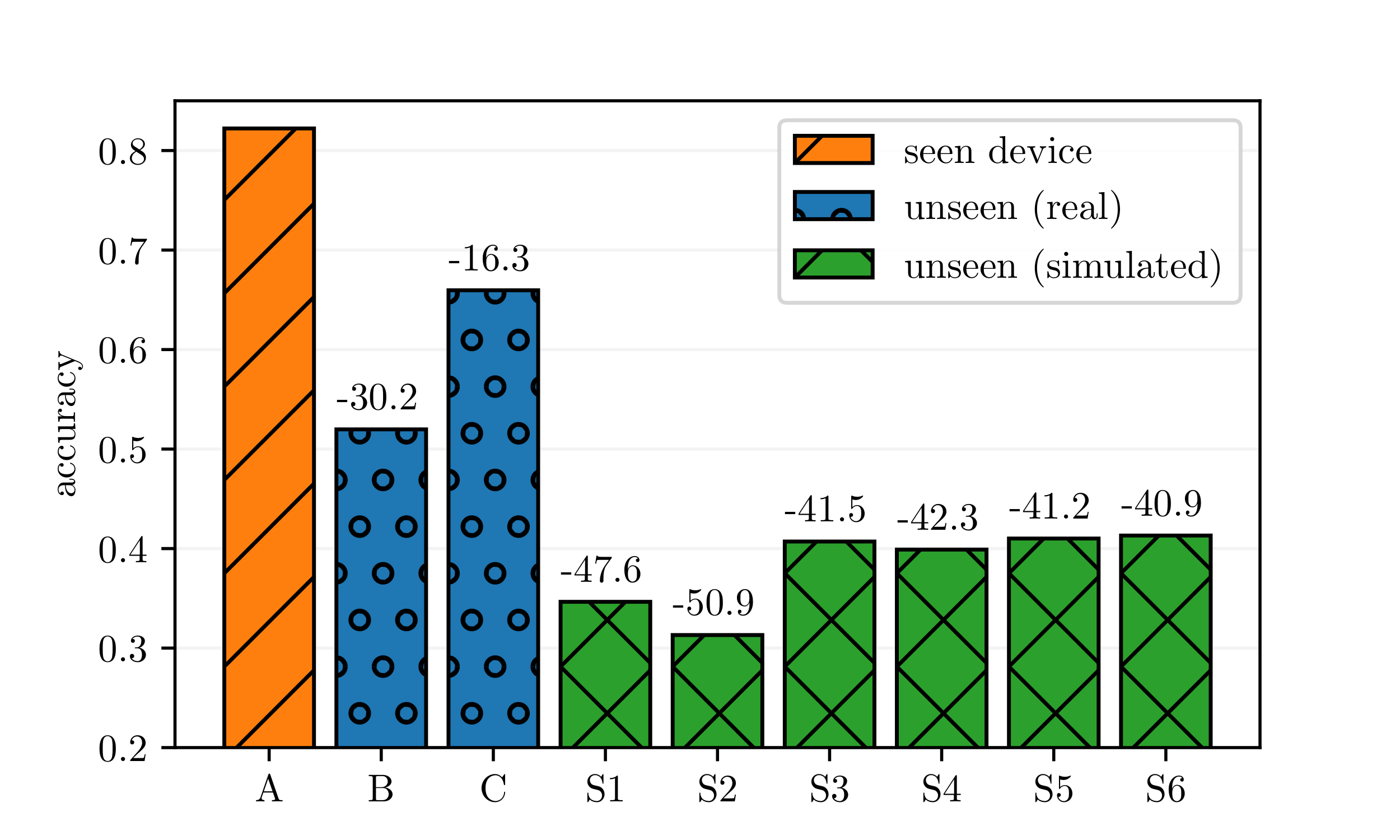}
    \caption{Acoustic scene classification performance of a vanilla audio spectrogram transformer on the TAU Urban Acoustic Scenes data set \cite{dcase2021} grouped by recording device. Classification accuracy degrades significantly for recording devices that are not present in the training set.}
    \label{figure:asc_on_unseen}
\end{figure}

Deep neural networks have become state-of-the-art tools for audio-related signal-processing tasks, such as acoustic scene classification, audio tagging, and sound event detection \cite{ast,passt,beats}. 
These models are known to generalize well if the recording conditions during training and later during inference remain the same; however, the generalization degrades when there is a distribution mismatch between the training and the testing data. 
Figure \ref{figure:asc_on_unseen} illustrates this problem on an Acoustic Scene Classification (ASC) task: The performance of an Audio Spectrogram Transformer (AST) \cite{ast} finetuned on audios recorded with one specific recording device is significantly worse on audios from unseen recording devices; domain adaptation techniques are required to alleviate the domain discrepancy.
In this work, we study this \textit{domain adaptation problem} in the context of audio classification with mismatching recording devices using ASTs. \\

\begin{figure*}[t]
  \centering
    \includegraphics[width=1\textwidth]{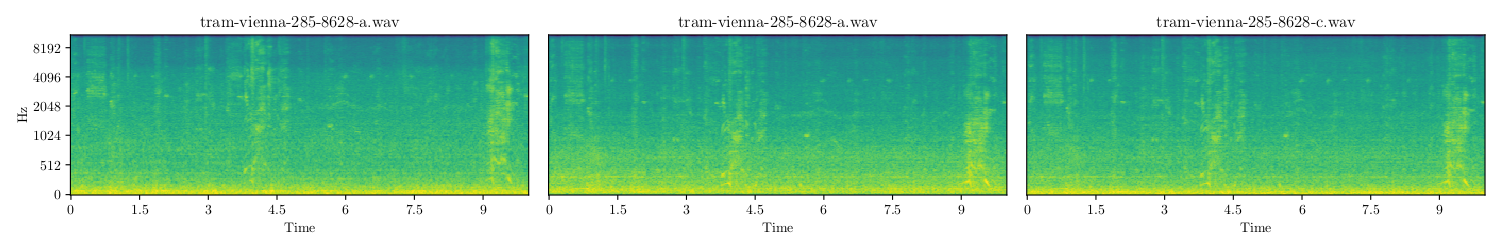}
    \caption{From left to right: Time-aligned recordings from devices A (Soundman OKM II Klassik/Studio A3 Microphone \& Zoom F8 Recorder), B (Samsung Galaxy S7), and C (iPhone SE) from the TUA Urban Acoustic Scenes data set 2019 \cite{dcase2021}. While the content is similar, the frequency magnitudes along the time dimension differ among the devices.}
    \label{figure:parallel_recordings}
    \vspace{-0.4cm}
\end{figure*}

\begin{figure*}[t]
  \centering
    \includegraphics[width=1\textwidth]{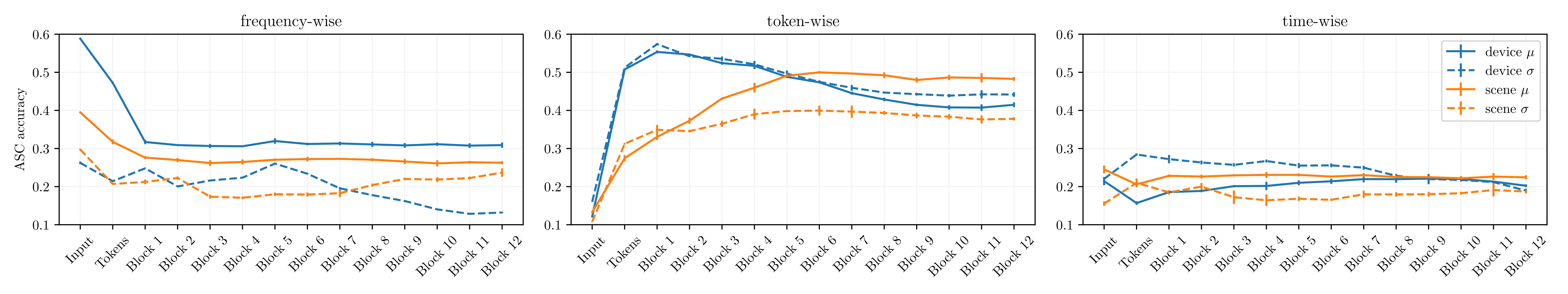}
    \caption{Classification accuracy of a random forest classifier on frequency-, channel-, and time-wise mean and standard deviation at different network depth in a self-supervised pretrained audio spectrogram transformer.}
    \label{figure:transformer_embeddings}
    \vspace{-0.4cm}
\end{figure*}

Previous investigations into the feature maps produced by Convolutional Neural Networks (CNNs) \cite{cnns} in the image domain \cite{batch_instance, meta_batch_instance, two_at_once} determined that visual domain characteristics (such as image style) are largely encoded in the channel dimensions of hidden representations. Channel-wise normalizations of the feature maps (Instance Norm \cite{instance_norm}) proved effective in alleviating the domain discrepancy.
However, unlike images, which encode spatial dimensions along both height and width of the input, height and width in visual audio representations encode frequency and time. 
While the spatial dimensions in images only carry little domain-relevant information, the frequency dimension in audio spectrograms carries important domain characteristics, such as information about the recording device. This is illustrated in Figure \ref{figure:parallel_recordings}, which shows a side-by-side comparison of parallel audio recordings taken with three different recording devices.
All three spectrograms show similar textures; their main difference is in the frequency bins' magnitudes. 
This consequently affects how frequency information is spread in the hidden representations: Kim et al. \cite{rfn} demonstrated that recording-device characteristics are not only present in the channel-wise statistics but also in the frequency-wise statistics of CNN hidden layer activations. To reduce the domain gap in audio CNNs, they introduced a combination of layer normalization \cite{layer_norm} and frequency normalization called Relaxed Instances Frequency-wise Normalization (RFN) for inputs and hidden activations. \newpage

Audio Spectrogram Transformers (ASTs) have recently become the new dominant model architecture for audio recognition tasks, outperforming CNNs on a broad variety of tasks (tagging, classification, etc.) and acoustic domains (speech, music, general sounds) \cite{ast, passt, beats}.
Our study sets out to investigate the usefulness of frequency normalizations for domain generalization in ASTs. CNNs use convolutions, which are local operations that only consider a limited context, while the self-attention layers in transformers operate on the whole spectrogram. 
This raises the question of how recording device-specific information is dispersed in hidden activations of transformers and whether similar frequency-wise normalization strategies as introduced for CNNs can be employed to encourage recording device invariance.
To address this question, we examined the hidden layer activations of ASTs to understand how and to what degree recording-device-specific information is encoded in their representations. Based on the observations of this initial investigation, we introduce a relaxed frequency-wise spectrogram centering operation that considerably improves the performance of ASTs on unseen recording devices. In the experiment section, we demonstrate that this simple method compares favorably to previously introduced methods for CNNs that normalize both the input and hidden activations.

\section{Analysis of AST Representations}

We focus on self-attention-based models and investigate where and to what degree recording-device characteristics are encoded in their hidden layer activations. To this end, we use an AST that was pre-trained on AudioSet in a self-supervised manner, to embed the TAU Urban Acoustic Scenes data set \cite{dcase2021} (see Section \ref{MSM} for details). Unlike CNNs, Spectrogram transformers process their input by cutting two-dimensional spectrograms into patches, converting them to one-dimensional tokens, and flattening the result to obtain a sequence. In order to restore the time and frequency structure from this sequential representation for analysis, we keep track of the time and frequency position of each token and reshape the activations after each self-attention block into a three-dimensional tensor $x \in \mathbb{R}^{D \times F \times T}$ where $D$, $F$ and $T$ are the sizes of the token, frequency and time dimensions, respectively. To quantify the importance of each dimension, we train a random forest classifier to predict the recording device or acoustic scene labels from the activations' dimension-wise mean and standard deviation statistics. We report the classification accuracy in Figure \ref{figure:transformer_embeddings}. We make three key observations:
\begin{itemize}

\item First, similarly to audio CNNs \cite{rfn}, ASTs concentrate recording device information in the frequency and token dimensions; time-wise statistics are less predictive (compare the left two plots to the rightmost one in Fig. \ref{figure:transformer_embeddings}). However, unlike in CNNs, the amount of device information in the frequency statistics drops significantly after the first self-attention block. We hypothesize that this is due to the global nature of self-attention, which draws information from the whole sequence rather than just from a local context, and conclude that frequency-wise normalization of hidden representations after the first self-attention block would only have a minor impact.

\item Next, we note that hidden layers' frequency- and token-wise statistics are predictive of both the recording device and the acoustic scene class, indicating that normalizations along the token dimension might also have a negative impact on the acoustic scene target prediction.

\item Finally, we observe that the mean over the frequency dimension predicts the recording device and the acoustic scene better than the standard deviation. Therefore, we conjecture that centering the spectrogram instead of both centering and whitening will be sufficient for better device generalization. 

\end{itemize}

\section{Relaxed Instance Frequency-Wise Centering}
Based on our previous analysis of the hidden-layer activations in transformers, we suspect that most of the recording device characteristics can be removed by centering the input spectrogram along the time dimension. We, therefore, define a frequency-wise centering operation for inputs or hidden layer representations $x \in \mathbb{R}^{D \times F \times T}$:
$$ FC(x) := x_{:, :, :} - \frac{1}{D \cdot T}\sum_{i=1}^{D} \sum_{j=1}^{T}  x_{i, :, j} $$
For mono-channel spectrograms, the token dimension $D$ reduces to a single channel. While this formulation is general enough to be used in arbitrary hidden layer activations, our previous analysis suggested that applying it after the first self-attention block will have little effect. We, thus, only apply it to the input spectrogram, and we evaluate this decision in the experiment section.
 Furthermore, since our previous analysis showed that the frequency-wise means are also correlated to the prediction target, we use a Softened version of Frequency-wise Centering (SFC) which is simply a convex combination of the original activations and the frequency-wise centered activations:
$$SFC(x) = \lambda \cdot FC(x) +(1 - \lambda) \cdot x $$
$\lambda$ controls the trade-off between the original and centered activations.

\section{Experiments}

\subsection{Data Set}

We use the TAU Urban Acoustic Scenes 2020 data set \cite{dcase2021} for our experiments as it provides prediction targets (ten acoustic scene locations) and recording device information which we need to estimate the device generalization performance. We train our system on the $10,215$ $10$-second snippets recorded with device A, and evaluate the device generalization on the $2,970$ segments in the evaluation set, which were recorded with real devices A, B, C, and simulated devices S1-S6. The samples in the evaluation set are class and recording-device balanced, so we use accuracy to measure task performance. We would like to refer the reader to the data set description \cite{dcase2021} for specifics on the recording procedure and creation of simulated devices.

\subsection{Features Computation}
We convert the audio recordings into Log-Mel spectrograms by computing a $1024$-point Short Time Fourier Transform using $40$ms windows and a hop size of $10$ms. We apply a Mel-scaled filter bank with $80$ filters for frequencies between $0$ and $8000$Hz and convert the result to $\textrm{dB}$ by applying a log transformation. The samples are normalized during training by subtracting the training set mean and dividing by the training set standard deviation. 

\subsection{Pretraining with Masked Spectrogram Modeling} \label{MSM}
Audio Spectrogram Transformers require a lot of training data to generalize; we, therefore, employ a self-supervised pre-training strategy called Masked Spectrogram Modeling (MSM) \cite{ssast, MSP, MSM} to alleviate this problem. In MSM, random parts of the spectrogram are removed, and the model is trained to predict the masked parts based on the available context. By doing so, the model is encouraged to learn a meaningful high-level representation that captures the underlying structure of sounds.
Our MSM framework follows an encoder-decoder architecture and we choose a standard ViT architecture \cite{vit} with $12$ self-attention blocks, $12$ encoder heads, and a token dimensionality of $768$ for the encoder. The decoder is smaller with only $4$ self-attention blocks, $6$ attention heads, and a dimensionality of $384$. Spectrograms are processed in the following manner: The mono-channel input spectrogram $x \in \mathbb{R}^{1 \times F \times T}$ is randomly shortened to $6$ seconds to reduce training time and cut into non-overlapping $16 \times 16$ patches. Each patch is converted into a one-dimensional token of size ${D}$ using a linear transformation. Separate learnable positional encodings for time and frequency dimensions are added to the tokens. We randomly drop 75\% of the input tokens and feed the resulting sequence into a self-attention-based encoder to obtain a high-level representation. The previously dropped tokens are then replaced in the encoded sequence with placeholder tokens, positional encoding is added, and the result is fed through the self-attention-based decoder. A linear projection head is used to generate spectrogram patches from the decoder output tokens. The model is trained to generate the removed parts of the input spectrogram, which is optimized by minimizing the mean squared error.
We train the model for $100$ epochs on AudioSet \cite{audioset} with the Adam optimizer ($\beta_1 = 0.9$, $\beta_2=0.95$) \cite{adam}, a batch size of $2048$ and a weight decay of $0.05$. The learning rate is linearly increased from $10^{-6}$ to $3\cdot 10^{-4}$ in the first $20$ epochs and decayed with a cosine schedule afterward. We validate  the performance of the embedding model on the HearEval benchmark \cite{hear} in Table \ref{tab:hear} and find the results on par with comparable models.

\begin{table}[h]
\caption{Hear Eval environmental sound tasks benchmark \cite{hear} results of our pre-trained model compared to \cite{MSM}.}
\centering
\begin{tabular}{@{}lllll@{}}
\toprule
     & ESC-50 \cite{esc} & FSD50K \cite{fsdk} & Gunshot \cite{gunshot} \\ \midrule
MSM \cite{MSM}  &   85.6        &   52.2     &   96.4      \\
ours &   86.8       &    52.4    &     91.4    \\ \bottomrule
\end{tabular}
\label{tab:hear}
\end{table}

\begin{table*}[t]
\caption{Device-wise comparison of ASC performance of our methods to the baselines. Columns PT and HN indicate pre-taining with centering and whether the normalization was used on hidden activations, respectively.}
\centering
\begin{tabular}{@{}llll|ll|l|llllllllll@{}}
\toprule
    & $\lambda$ & PT  & HN & \multicolumn{1}{c}{all} &                               & \multicolumn{1}{c|}{seen} & \multicolumn{8}{c}{unseen}                                                                                                                                                                                &                         &                              \\ \midrule
Dev. ID  &  &  &  & \multicolumn{1}{c}{}    & \multicolumn{1}{c|}{$\Delta$} & \multicolumn{1}{c|}{A}    & \multicolumn{1}{c}{B}       & \multicolumn{1}{c}{C} & \multicolumn{1}{c}{S1} & \multicolumn{1}{c}{S2} & \multicolumn{1}{c}{S3} & \multicolumn{1}{c}{S4} & \multicolumn{1}{c}{S5} & \multicolumn{1}{c}{S6} & \multicolumn{1}{c}{avg} & \multicolumn{1}{c}{$\Delta$} \\ \midrule
Baseline & 0.0 & \xmark & \xmark & $47.7 \pm 1.6$ & $+ 0$ & $82.2$ & $52.0$ & $66.0$ & $34.7$ & $31.3$ & $40.7$ & $39.9$ & $41.0$ & $41.3$ & $43.4$ & $+0$ \\
GFN  & $-$ & \xmark &\xmark & $42.1 \pm 1.0$ & $-5.6$ & $80.7$ & $46.8$ & $65.8$ & $31.7$ & $25.9$ & $32.3$ & $32.8$ & $31.3$ & $31.5$ & $37.3$ & $+5.6$\\ 
RFN \cite{rfn} & $0.5$ & \xmark & \cmark & $49.9 \pm 0.1$ & $+2.3$ & $57.7$ & $51.4$ & $51.1$ & $50.3$ & $49.2$ & $49.0$ & $49.9$ & $47.6$ & $43.1$ & $48.9$ & $+5.6$\\ 
RFN & $0.9$ & \xmark & \xmark & $60.7 \pm 0.3$ & $+13.0$ & $73.0$ & $61.2$ & $68.1$ & $57.3$ & $54.6$ & $60.1$ & $58.2$ & $58.8$ & $54.9$ & $59.1$ & $+15.8$ \\ \midrule
SFC & 0.4 & \xmark &\xmark & $50.4 \pm 0.3$ & $+2.8$ & $\mathbf{84.0}$ & $55.3$ &  $67.8$ & $43.1$ & $31.7$ & $41.1$ & $42.7$ & $44.3$ & $43.7$ & $46.2$ & $+2.9$ \\ 
SFC & $0.9$ & \xmark &\xmark & $\mathbf{62.9 \pm 0.6}$ & $\mathbf{+15.2}$ & $73.8$ & $65.2$ & $\mathbf{71.0}$ & $\mathbf{60.8}$ & $\mathbf{54.7}$ & $\mathbf{63.9}$ & $57.0$ & $59.8$ & $\mathbf{59.7}$ & $\mathbf{61.5}$ & $ \mathbf{+ 18.2}$ \\
SFC & $1.0$ & \xmark & \xmark & $62.1 \pm 0.1$ &  $+14.4$ & $72.7$ & $\mathbf{65.8}$ & $68.7$ & $58.1$ & $54.4$ & $62.5$ & $\mathbf{58.3}$ & $\mathbf{60.2}$ & $57.8$ & $60.7$ & $+17.4$ \\ \midrule
SFC & $0.9$ & \cmark & \xmark & $55.1 \pm 1.0$ &  $+7.5$ & $79.8$ & $58.2$ & $68.8$ & $47.8$ & $44.8$ & $49.0$ & $46.$ & $52.1$ & $49.1$ & $52.0$ & $+8.7$ \\ 
\bottomrule
\end{tabular}

\label{tab:main_result}
\end{table*}

\subsection{Fine Tuning}
For fine-tuning on the TAU Urban Acoustic Scenes data set \cite{dcase2021}, we discard the decoder and linearly interpolate the positional encoding to handle spectrograms of $10$ seconds in length. The spectrogram processing is similar to the procedure described above, but only $40\%$ of the input tokens are dropped \cite{passt}. In addition, we use gain augmentation with $\pm 7db$, and MixUp \cite{mixup} on the raw audio waveforms ($p=0.5$ and $\alpha=2.$), Mixup on the Log-Mel spectrogram ($p=1.0$ and $\alpha=0.3$) and SpecAugment \cite{specaugment} with maximum frequency and time width of $30$ and $192$, respectively. The model is trained for $100$ epochs using Adam optimizer with weight decay $10^{-4}$ and a batch size of $64$. The learning rate is linearly increased to $5\cdot 10^{-5}$ during the first five epochs and decayed using the cosine decay rule.

\subsection{Baselines}
We compare our method to a weak baseline without a domain adaptation strategy ($\lambda = 0$) and three strong baselines that employ frequency-wise normalizations: first, Global Frequency Normalization (GFN), which uses frequency normalization statistics computed over the training data set. Second, Relaxed Instance Frequency-wise Normalization (RFN) \cite{rfn}, which was originally introduced for CNNs and uses a combination of layer normalization \cite{layer_norm} and instance frequency-wise normalization of both the input and hidden activations. And thirdly, a modified version of RFN that only normalizes the input spectrogram.

\section{Results}
We run three repetitions of grid searches over $\lambda$ in $\{0.0, 0.1, \dots, 1.0\}$ for SFC and both variants of RFN and report the best average performance together with the best value for $\lambda$ in Table \ref{tab:main_result}. 
Compared to the weak baseline with $\lambda = 0$, SFC with $\lambda=0.9$ increases the performance on unseen recording devices by $18.15$ pp. However, we also notice a considerable drop in performance on recording device A, namely by $8.3$ pp. With $\lambda=0.4$, the performance across all recording devices increases, but only moderately.
For the original version of RFN, we obtain the best results with $\lambda=0.5$. However, the frequency-wise normalizations of hidden layer activations seem to have a negative impact, which can be seen by the large performance increase when using the normalization on the input spectrogram only. We further observe a slight performance gap between RFN and SFC which we attribute to the layer normalization used in RFN.

\subsection{Impact of $\lambda$}
\begin{figure}[t]
  \centering
    \includegraphics[width=1\linewidth]{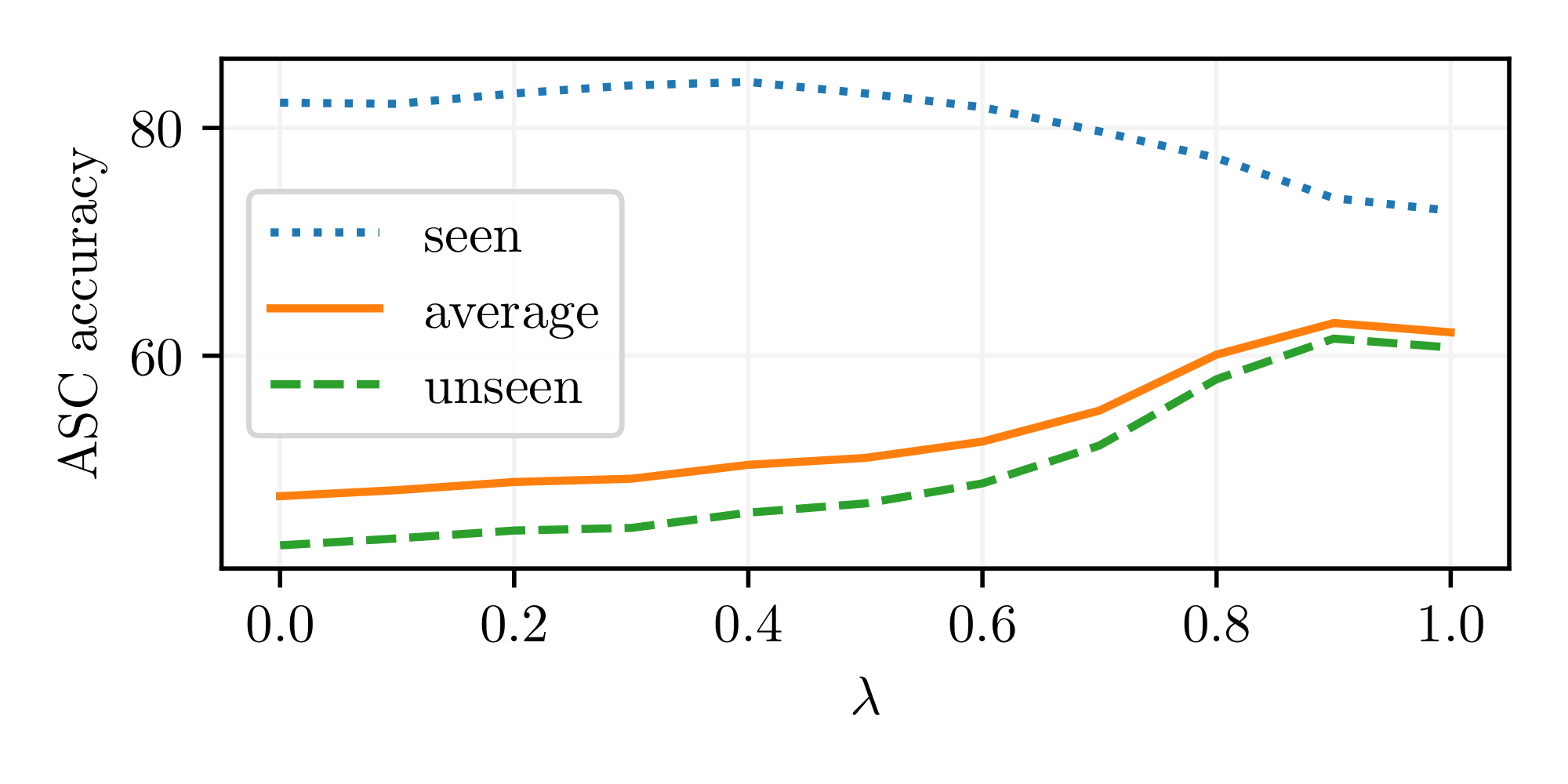}
    \caption{Performance of frequency-wise centering on seen and unseen recording devices for different degrees of centering ($\lambda=0$ means no centering.)}
    \label{figure:lambda}
\end{figure}

We further investigate the effect of relaxation by plotting the performance on seen and unseen devices over different degrees of relaxation in Figure \ref{figure:lambda}. Lower values of $\lambda$ lead to small improvements on the original and the unseen devices. With even higher values for $\lambda$, the accuracy on unseen devices increases even further and peaks at $\lambda = 0.9$. However, this comes at the price of a decreased performance on the seen devices, which drops significantly for $\lambda > 0.5$. This suggests that $\lambda$ needs to be fine-tuned based on the expected prior probability of seen and unseen recording devices.

\subsection{Centering of Hidden Representations}

We also investigate the effect of hidden layer centering. To this end, we retrained with two modifications: frequency-wise centering in all hidden layers and frequency-wise centering after the linear transformation from patches to tokens only. We search for the best value of $\lambda$ as above and report the results in Table \ref{tab:hidden}. While we see an improvement on the unseen devices over the baseline for all methods, normalizing only the input spectrogram yields the highest improvements on the unseen devices, which is in correspondence with our previous analysis of the hidden layer activations.

\begin{table}[h]

\caption{First section: Effect of applying the centering operation on the spectrogram, the input tokens, or the intermediate representations between self-attention blocks. Second section: Impact of centering and whitening instead of centering only.}
\centering
\begin{tabular}{@{}l|lll|l|lll@{}}
\toprule
    $\lambda$ & input & first & linear  & std & all & seen & unsee \\ \midrule
    $0.0$ & \xmark & \xmark & \xmark  & \xmark  &    $47.66$       &   $\mathbf{82.22}$     &   $43.35$      \\
    $0.9$ &\cmark &  \xmark & \xmark  & \xmark  &     $\mathbf{62.87}$       &   $73.84$     &   $\mathbf{61.50}$      \\
    $1.0$ &\xmark &  \cmark & \xmark  & \xmark  &     $57.29$       &   $69.70$    &   $55.83$    \\ 
    $0.4$ & \cmark &  \cmark & \cmark  & \xmark  &    $54.04$       &   $67.68$    &   $52.34$    \\ \midrule
    $0.8$ & \cmark &  \xmark & \xmark  &  \cmark  &    $61.30$       &   $73.33$    &   $59.80$    \\ 
    \bottomrule
\end{tabular}

\label{tab:hidden}
\end{table}

\subsection{Centering vs. Normalization}
Our initial investigation into transformer activations indicated that frequency-wise centering of the spectrogram might be sufficient to remove device information. We validate our decision by comparing our methods to a slight variation of the proposed method that centers and whitens the spectrogram frequency-wise. We grid search for the best choice of $\lambda$ in the same range as above and report the result in Table \ref{tab:hidden}. Additional whitening of the spectrogram did not increase the performance on seen and unseen devices; these results seem to suggest that the whitening step is redundant. 

\subsection{Pre-training with Normalization}
We further test whether pre-training the AST with normalized spectrograms could further improve the results. To this end, we repeat the pre-training procedure, but use SFC ($\lambda=0.9$) already during pre-training this time. The results are reported in the last row of Table \ref{tab:main_result}. Although we still observe a noticeable gain across all unseen recording devices, the performance gain is not as large as without normalization during pre-training. This finding is rather disappointing, but we also notice that the performance on the seen device does not degrade as significantly as without normalization during pre-training. We hypothesize that a different choice for $\lambda$ could lead to better results, but have not run large-scale experiments to pinpoint the optimal lambda value. 

\section{Conclusion}

 The main goal of this study was to investigate frequency-wise normalization strategies to encourage recording device invariance in ASTs. Our initial investigation into their hidden layer activations suggested that most of the recording device information is transferred from the frequency dimension into the token dimension after the first self-attention block. We designed a frequency-wise centering operation for spectrograms to remove recording device characteristics at an early stage, which greatly improved the target prediction accuracy on unseen recording devices. A natural line of progression of this work would be to look into self-supervised training methods that incorporate strategies for recording device invariant learning during pre-training.

\bibliographystyle{ieeetr}
\bibliography{refs}
 
\end{document}